\RequirePackage{fix-cm}

\documentclass[twocolumn,epjc3]{svjour3}
\smartqed  
\RequirePackage{graphicx}

\journalname{Eur. Phys. J. C}
\begin{document}

\title{Magnetized particle motion around magnetized Schwarzschild-MOG black hole}

\author{Kamoliddin~Haydarov  \thanksref{e1,addr1} \and Javlon~Rayimbaev \thanksref{e2, addr2, addr3, addr4} \and Ahmadjon~Abdujabbarov \thanksref{e3, addr2, addr3, addr4, addr5, addr6} \and Satimbay~Palvanov \thanksref{e4, addr3} \and Dilfuza~Begmatova \thanksref{e5, addr3}}
\thankstext{e1}{haydarovk378@gmail.com}
\thankstext{e2}{javlon@astrin.uz}
\thankstext{e3}{ahmadjon@astrin.uz}
\thankstext{e4}{satimbay@yandex.ru}
\thankstext{e5}{dilfuza.fizic@mail.ru}

\institute{Tashkent University of Information Technologies named after Muhammad al Khwarezmi, Amir Temur 108, Tashkent 100014, Uzbekistan \label{addr1}\and Ulugh Beg Astronomical Institute, Astronomicheskaya 33, Tashkent 100052, Uzbekistan \label{addr2} \and National University of Uzbekistan, Tashkent, Uzbekistan \label{addr3} \and Institute of Nuclear Physics, Ulugbek 1, Tashkent 100214, Uzbekistan \label{addr4} \and Shanghai Astronomical Observatory, 80 Nandan Road, Shanghai 200030, P. R. China \label{addr5} \and Tashkent Institute of Irrigation and Agricultural Mechanization Engineers, Kori Niyoziy, 39, Tashkent 100000, Uzbekistan \label{addr6}}

\date{Received: date / Accepted: date}

\maketitle

\begin{abstract}
In this paper, we have presented the studies of the motion of magnetized particles and energetic processes around Schwarzschild black holes in modified gravity (MOG). The study of circular stable orbits shows that orbits of magnetized particles can not be stable for the values of magnetic coupling parameter $\beta \geq 1$. It was also shown that the range of stable circular orbits increases with the increase of both MOG and magnetic coupling parameters, while the effects of magnetic interaction stronger than the gravity. It was obtained that the increase of the MOG parameter causes the increase of center-of-mass energy collision of magnetized particles. Moreover, we have analyzed how to mimic the magnetic interaction with the spin of Kerr and Schwarzschild-MOG black holes. We have obtained that the magnetic coupling parameter can mimic the spin parameter $a \leq 0.15$ ($a \leq 0.28$) giving the same radius of innermost contour(co)-rotating orbits at the values of the parameter $\beta \in (-1,1)$ and the MOG parameter in the range $\alpha \in (-0.17,0.28)$ while the MOG parameter  $\alpha \in (-0.7, 0.9)$ mimics spin parameter of the black hole with the range $|a| \in (0,1)$.


\end{abstract}

\section{Introduction}

Detection of the dark energy and the dark matter in the Universe at the end of the 20th century changed our imaginations about Nature. The gravitational field described by the general relativity can not fully explain the nature of the dark energy and the dark matter. Among the different ways of explaining their nature the modified gravity (MOG) proposed in~\cite{Moffat06} became one of the most promising model. The theory proposed in~\cite{Moffat06} considers the scalar and massive vector field and usually called scalar-tensor-vector gravity (STVG). The motivation of introducing MOG is that being the classical field theory the general relativity breaks down at short length scales. In order to consider the quantum effects one needs the modification of the theory and MOG is one of the way to modify the general relativity. Introduced massive vector field with the source charge $Q=\sqrt{\alpha G}M$, where $G$ is the gravitational constant, $M$ is the mass of the central object, and $\alpha$ is the new coupling parameter, causes repulsive force, and becomes significant at the quantum level.

The non-rotating and rotating black hole solutions within MOG theory have been obtained in~\cite{Moffat15} and called Schwarzschil-MOG and Kerr-MOG black hles, respectively.
%
%
The test particle motion around Schwarzschild-MOG black hole has been studied in~\cite{Hussain15}, and it was shown the orbits become more stable due to the presence of a vector field in STVG theory. The several interpretations of MOG theory through solar system tests~\cite{Moffat06}, galaxy rotation curve~\cite{Moffat13,Moffat15a}, through X-ray observations~\cite{Moffat14},
 black hole shadow~\cite{Moffat15b,Moffat15}, the study of thermodynamics~\cite{Mureika16}, supernovae~\cite{Wondrak18}, gravitational lensing~\cite{Moffat09}, quasinormal modes~\cite{Manfredi18}. The thermodynamic properties of MOG theory have been studied in~\cite{Pradhan18}. Epicyclic frequencies in Kerr-MOG black hole discussed in~\cite{Pradhan19}. Quasi-periodic oscillations around Kerr-MOG black holes have been studied in~\cite{Kolos20}. 	Test particle dynamics near Kerr-MOG black hole have been considered in~\cite{Sharif17}. The gravitational instability in the Newtonian limit of MOG has been discussed in~\cite{Shojai17}.

A Magnetic field surrounded black hole is one of the useful tests of gravity theories. The electromagnetic field around Kerr black hole immersed in an external asymptotically uniform magnetic field has been studied in the pioneering work of Wald ~\cite{Wald74}.
Then numerous authors studied the properties of the electromagnetic field around black hole in the presence of external uniform and dipolar magnetic field~\cite{Kolos17,Kovar10,Kovar14,Aliev89,Aliev02,Aliev86,Frolov11,Frolov12,Shaymatov18,Oteev16,Toshmatov15d,Stuchlik14a,Abdujabbarov14,Abdujabbarov10,Abdujabbarov11a,Abdujabbarov11,Abdujabbarov08,Karas12a,Stuchlik16} and neutron stars ~\cite{Rayimbaev2020MPLA,Turimov18b,Turimov17,Rayimbaev15,Rayimbaev2019IJMPD}.

In the presence of electromagnetic field around black hole, one may study the motion of particles with non-zero spin and magnetic dipole momentum.
In Ref.~\cite{deFelice} it was shown that around Schwarzschild black hole immersed in an external asymptotically uniform magnetic field the magnetized particles can move along stable non-geodesic, spatially circular equatorial orbits with the radius smaller than Innermost stable circular orbits (ISCO).
The study has been extended to the case of Kerr black hole in Ref.~\cite{defelice2004}.

One of the authors of this paper has studied the magnetized particle motion around non-Schwarzschild black hole in the presence of a magnetic field ~\cite{Rayimbaev16}. Other our study is devoted to acceleration of magnetized particle around a rotating black hole in quintessence~\cite{Oteev16}. One of the autors of this paper has been involved to study the high energy collision of magnetized particles around a Ho\v{r}ava-Lifshitz black hole~\cite{Toshmatov15d}. Magnetized particle acceleration around a Schwarzschild black hole in a magnetic field has been analyzed in~\cite{Abdujabbarov14}. One may find the analysis of magnetized particle motion around braneworld black hole in~\cite{Rahimov11a,Rahimov11}. The magnetized particle motion in conformal gravity has been study in~\cite{Haydarov20}.

The Penrose process~\cite{Penrose69a}, Blandford-Znajeck mechanism~\cite{Blandford1977}, Magnetic Penrose process~\cite{Dhurandhar83,Dhurandhar84,Dhurandhar84b,wagh85} , and particle acceleration mechanism (BSW)~\cite{Banados09} are considered as a toy model of different energetic processes around compact objects in astrophysics. For the review of energetic processes in different models of gravity we refer the Reader to the following references~\cite{Stuchlik14a,Abdujabbarov14,Abdujabbarov10,Abdujabbarov11a,Abdujabbarov08,Abdujabbarov13a,Abdujabbarov13b,Narzilloev19}.

This work is devoted to studying the effect MOG to magnetized particle motion around a black hole and acceleration process of this type of particles. The paper is organized as follows: The Sect.~\ref{chapter1} is devoted to study the electromagnetic field and magnetized particle motion around Schwarzschild-MOG black hole. The magnetized particle acceleration near the Schwarzschil-MOG black hole has been studied in Sect.~\ref{magaccel}. We consider some astrophysical applications of our results in Sect.~\ref{astroapll}. We summarize our results in Sect.~\ref{Summary}.

We use signature $(-,+,+,+)$ for the space-time and geometrized unit system $G_{\rm N}= c = 1$. The Latin indices run from $1$ to $3$ and the Greek ones from $0$ to $3$.

\section{Magnetized particle motion around Schwarzschild-MOG black holes in magnetic field \label{chapter1}}

The spacetime metric around Schwarzschild black holes in modified gravity can be described as~\cite{Moffat15}:
\begin{eqnarray}
ds^2=-fdt^2+f^{-1}dr^2+d\Omega^2\ ,
\end{eqnarray}
where
\begin{equation}
f=1-\frac{2(1+\alpha)M}{r}+\frac{\alpha(1+\alpha)M^2}{r^2}\ ,
\end{equation}
and $\alpha$ is coupling parameter of MOG gravity.

Consider the Schwarzschild-MOG black hole immersed in an asymptotically uniform magnetic field. The electromagnetic four-potential can be found using the Wald method~\cite{Wald74} and expressed as:
\begin{eqnarray}
A_{\phi} & = & \frac{1}{2}B_0r^2\sin\theta,
\end{eqnarray}
where $B_0$ is external uniform magnetic field. The non zero components of the electromagnetic tensor can be easily calculated using the definition $F_{\mu\nu}=A_{\nu,\mu}-A_{\mu,\nu}$ and have the following form
\begin{eqnarray}\label{FFFF}
F_{r \phi}&=&B_0r\sin^2\theta\ ,
 \\
 F_{\theta \phi}&=&B_0r^2\sin\theta \cos\theta\ .
\end{eqnarray}
The nonzero orthonormal components of magnetic field in the rest frame of the comoving observer have the following form
\begin{equation}
    B^{\hat{r}}=B_0 \cos\theta, \\ \qquad B^{\hat{\theta}}=\sqrt{f}B_0\sin \theta
\end{equation}

Now we construct the equations of motion of magnetized particles around Schwarzschild black hole immersed in the external asymptotically uniform magnetic field in MOG theory. The Hamilton-Jacobi equation for magnetized particles can be expressed in the following form\cite{deFelice}
\begin{eqnarray}\label{HJ}
g^{\mu \nu}\frac{\partial {\cal S}}{\partial x^{\mu}} \frac{\partial {\cal S}}{\partial x^{\nu}}=-\Bigg(m-\frac{1}{2} D^{\mu \nu}F_{\mu \nu}\Bigg)^2\ ,
\end{eqnarray}
where $m$ is mass of the particle, $S$ is the actin for magnetized particle in the curved spacetime background, the product of polarization and electromagnetic field tensors $D^{\mu \nu}F_{\mu \nu}$ is responsible of the interaction between the external magnetic field and magnetized particles. The expression for the polarization tensor $D^{\mu \nu}$ corresponding to the magnetic moment of magnetized particles has the following form~\cite{deFelice}:
\begin{eqnarray} \label{Dab}
D^{\alpha \beta}=\eta^{\alpha \beta \sigma \nu}u_{\sigma}\mu_{\nu} , \qquad D^{\alpha \beta }u_{\beta}=0\ ,
\end{eqnarray}
where $\mu^{\nu}$ is the four-vector of magnetic dipole moment and $u^{\nu}$ is four-velocity of the particles in the rest frame of the fiducial comoving observer, being orthogonal to the magnetic moment. The electromagnetic field tensor $F_{\alpha \beta}$ can be expressed through electric $E_{\alpha}$ and magnetic $B^{\alpha}$ field components as
\begin{eqnarray}
F_{\alpha \beta}=u_{[\alpha}E_{\beta]}-\eta_{\alpha \beta \sigma \gamma}u^{\sigma}B^{\gamma}.
\end{eqnarray}

Using the condition given in equation (\ref{Dab}) one can easily calculate the interaction quantity $D\cdot F$ in the following form
\begin{eqnarray}\label{DF1}
 D^{\mu \nu}F_{\mu \nu}=2\mu^{\hat{\alpha}}B_{\hat{\alpha}}
 =2\mu B_0 {\cal L}[\lambda_{\hat{\alpha}}]\ ,
 \end{eqnarray}
where $\mu=|\mu|=\sqrt{\mu_{\hat{i}}\mu^{\hat{i}}}=\mu$ is the module of the dipole magnetic moment of the magnetized particles and $L[\lambda_{\hat{\alpha}}]$ is the function of the coordinates and the other parameters of the spacetime around the black hole as well as  magnetic field defining the tetrad ${\lambda_{\hat{\alpha}}}$ attached to the comoving fiducial observer.

Here we study the orbital motion of magnetized particles
around the Schwarzschild-MOG black hole in the weak magnetic interaction approximation, in other word we dropped out higher orders of $\Big(D^{\mu \nu}F_{\mu \nu}\Big)^2\rightarrow{0}$ . The conserved quantities in the equatorial plane are the angular momentum $p_{\phi}= L$ and energy $p_t = -E$ of the particle. Now one can express the action of the magnetized particle in the form
\begin{eqnarray}
{\cal S}=-Et+L\phi +S_r\ ,
\end{eqnarray}
which can be used to seperate the variables in Hamilton-Jacobi equation. The equation of radial motion of the magnetized particles can be found in the following form
\begin{eqnarray}
\dot{r}^2={\cal{E}}^2-1-2V_{\rm eff}(r,\alpha,l,\beta)\ ,
\end{eqnarray}
where newly introduced effective potential of radial motion has the form:
\begin{eqnarray}
V_{\rm eff}(r;\alpha,l,\beta)=\frac{1}{2}\Bigg[f\left(1+\frac{l^2}{r^2}-\beta {\cal L}[\lambda_{\hat{\alpha}}]\right)-1\Bigg]\ ,
\end{eqnarray}
where $\beta = 2\mu B_0/m$ is the magnetic coupling parameter and $l=L/m$ is the specific angular momentum of the magnetized particle.

The condition for the circular orbits can be expressed as follows:
\begin{eqnarray}\label{circular}
\dot{r}=0  , \qquad \frac{\partial V_{\rm eff}(r;\alpha,l,\beta)}{\partial r}=0\ .
\end{eqnarray}
The first condition in~(\ref{circular}) allows one to find the possible values of the magnetic coupling parameter $\beta$ for circular orbits
\begin{eqnarray}\label{beta1}
\beta(r;l,{\cal E},\alpha)=\frac{1}{ {\cal L}[\lambda_{\hat{\sigma}}]}\Bigg(1+\frac{l^2}{r^2}-\frac{{\cal{E}}^2}{f}\Bigg)\ .
\end{eqnarray}
The second condition in~(\ref{circular}) gives us
\begin{eqnarray}
\frac{\partial V_{\rm eff}(r;\alpha,l,\beta)}{\partial r}=f {\cal L}[\lambda_{\hat{\alpha}}]\frac{\partial \beta (r;\alpha,l,\beta)}{\partial r}
\end{eqnarray}
We consider the particle at the equatorial plane with magnetic dipole moment perpendicular to the equatorial plane, thus the components of the external magnetic field measured by the observer in a frame of comoving with the particle take the following form
\begin{eqnarray}\label{Bcomp}
B_{\hat{r}}=B_{\hat{\phi}}=0\ , \qquad B_{\hat{\theta}}=B_0f\,e^{\Psi}\ .
\end{eqnarray}
Inserting Eq.(\ref{Bcomp}) into (\ref{DF1}) we can find the interaction part of the Eq.~(\ref{HJ})
\begin{eqnarray}\label{DF2}
  D \cdot F=2\mu B_0f\,e^{\Psi}\ ,
  \end{eqnarray}
where $e^{\Psi}=\left(f-\Omega^2 r^2\right)^{-\frac{1}{2}}$~\cite{deFelice} and $\Omega$ is angular momentum and has following form
\begin{eqnarray}
  \Omega=\frac{d\phi}{d t}=\frac{d\phi/d\tau}{d t/d\tau}=\frac{f}{r^2}\frac{l}{{\cal{E}}}\ .
\end{eqnarray}

Comparing Eq.(\ref{DF2}) with Eq.(\ref{DF1}) we get
\begin{eqnarray}
  {\cal L}[\lambda_{\hat{\sigma}}]=e^{\Psi}\, f\ .
\end{eqnarray}

Finally, the magnetic coupling parameter $\beta(r;l,{\cal E},\alpha)$ for stable circular orbits has the following form
\begin{eqnarray}\label{betafinal}
  \beta(r;l,{\cal E},\alpha)=\left(\frac{1}{f}-\frac{l^2}{{\cal E}^2 r^2}\right)^{1/2}\Bigg(1+\frac{l^2}{r^2}-\frac{{\cal E}^2}{f}\Bigg)\ .
\end{eqnarray}
Eq.~(\ref{betafinal}) implies that a magnetized particle with magnetic coupling parameter $\beta$ corresponds to circular stable orbit $r$ with the energy ${\cal E}$ and angular momentum $l$.
\begin{figure}[ht!]
  \centering
   \includegraphics[width=0.98\linewidth]{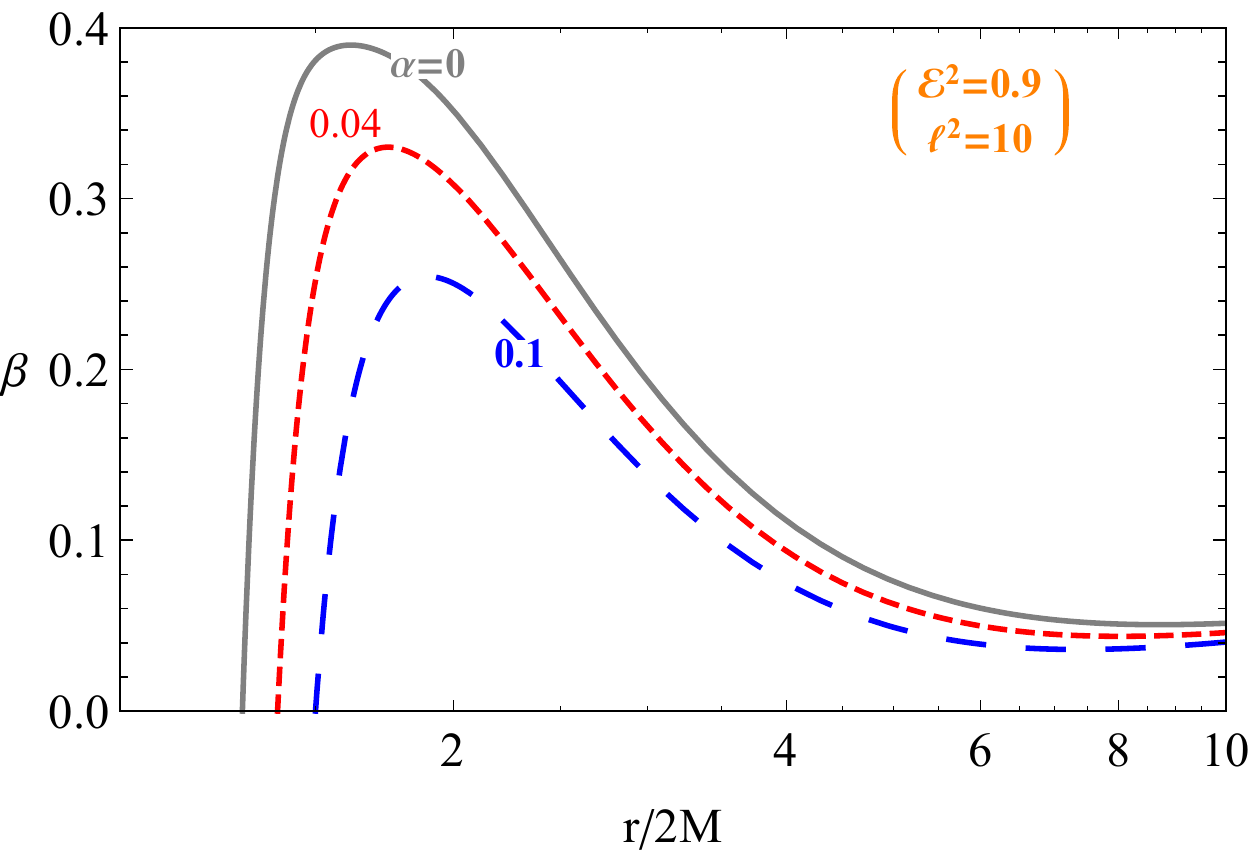}
   \includegraphics[width=0.98\linewidth]{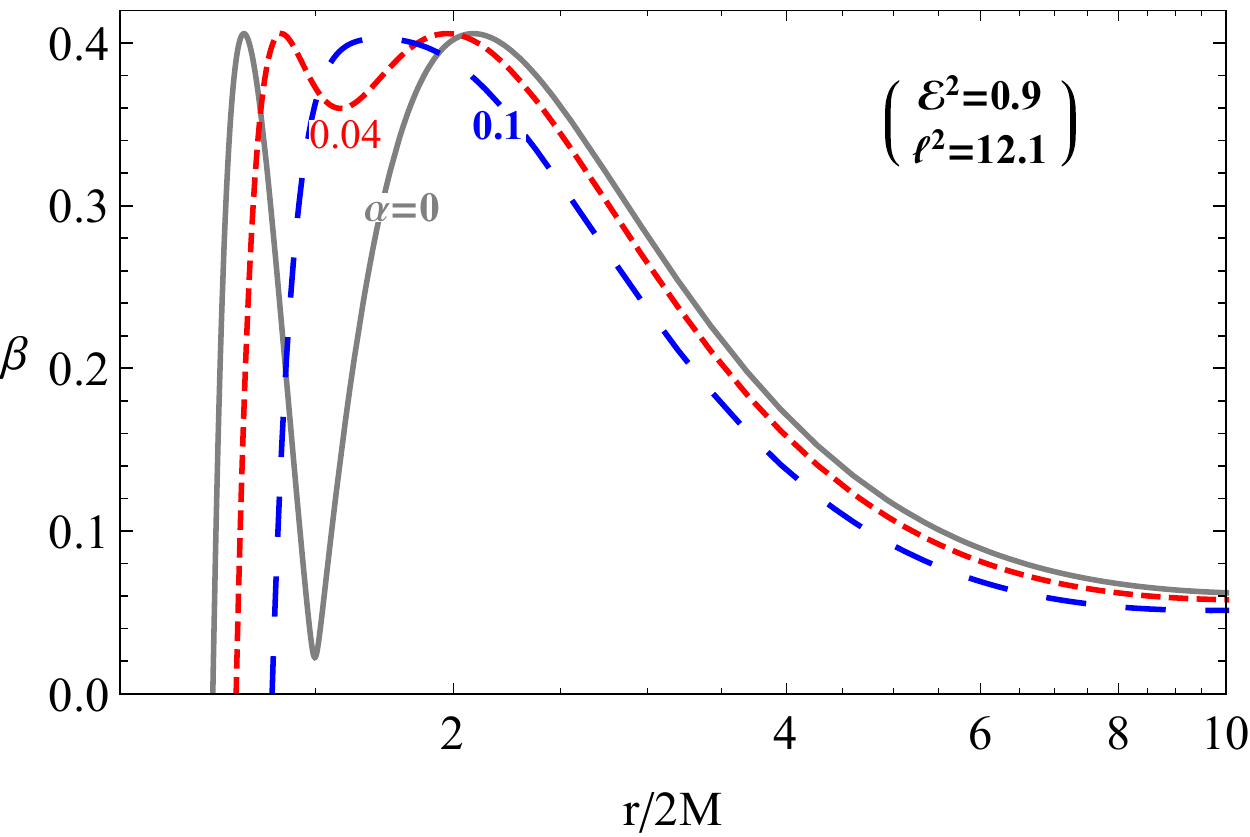}
 \caption{The radial dependence of magnetic coupling parameter for the different values of $\alpha$ parameter. The plots are taken for the value of the specific energy  ${\cal E}=\sqrt{0.9}$ and angular momentum $l=\sqrt{5}$ (top panel) and $l=\sqrt{6.074}$ (bottom panel). \label{betafig}}
\end{figure}

Fig.~\ref{betafig} shows the radial dependence of the magnetic coupling parameter $\beta$ for the different values of MOG parameter $\alpha$ for the fixed values of the specific energy and angular momentum. One can see from the top panel of Fig.~\ref{betafig} (when $l^2=10$) that the increase of the MOG parameter, $\alpha$, causes the decrease of the maximum value of the magnetized parameter. One may also see that  with the increase of the parameter, $\alpha$,  the loci where the parameter $\beta=0$ shifts to the observer at infinity. However, when $l^2>12$ (bottom panel of Fig.~\ref{betafig}), the maximum value of the magnetic coupling parameter $\beta$ does not  depend on the MOG parameter, $\alpha$. Moreover, the local minimum of the $\beta$ increases with the increase of $\alpha$ parameter and disappears for higher values of the parameter $\alpha$.

Now we analyze the values of the magnetic coupling parameter corresponding to the stable orbits of the magnetized particles. It can be found using following set of equations~\cite{deFelice,Rayimbaev16}:
\begin{eqnarray}\label{condbeta}
\beta =\beta(r,l,{\cal E},\alpha), \qquad \frac{\partial \beta(r,l,{\cal E},\alpha)}{\partial r}=0\ .
\end{eqnarray}
One can see~(\ref{condbeta}) contains two equations with five parameters of the particle ($\beta,r,l,{\cal E}$) and spacetime ($\alpha$), so its solution
can be parametrized in terms of any two of five independent variables. In order to solve the system of equations, it is better to use the magnetic coupling parameter and radius of the stable orbits $r$ as free parameters. First, we will find the specific energy ${\cal E}$ and the angular momentum $l$ of the magnetized particle as functions of radial coordinates and $\alpha$ parameter as:
\begin{eqnarray}\label{emin}
{\cal E}_{\rm min}^2=\frac{l^2 \left[\alpha  (\alpha +1)M^2-2 (\alpha +1) M r+r^2\right]}{r^2\sqrt{(\alpha +1) M  (r-\alpha M)}}\ .
\end{eqnarray}
The expression (\ref{emin}) corresponds to the  possible values of the specific energy of the magnetized particle at stable circular orbits.

\begin{figure}[ht!]
  \centering
   \includegraphics[width=0.98\linewidth]{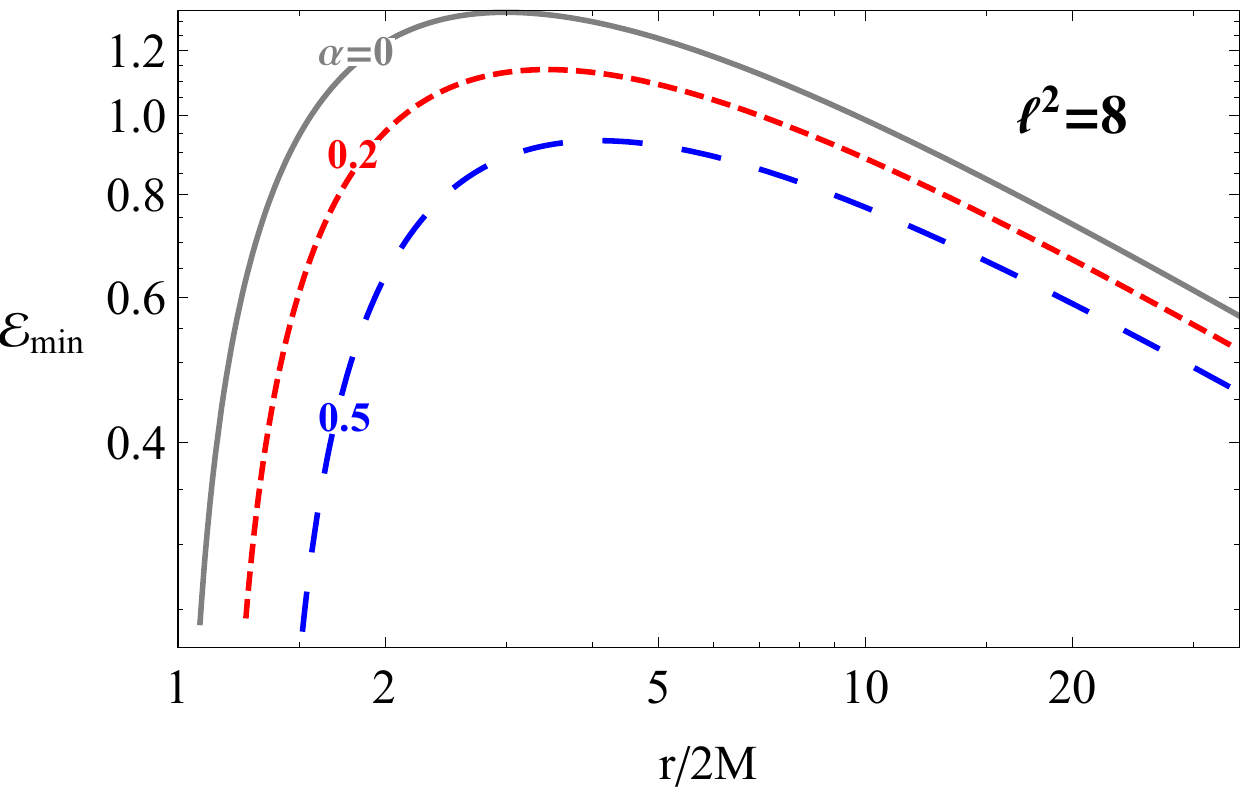}
    \caption{Radial ptofile of minimal energy of the magnetized particle. \label{eminfig}}
\end{figure}

Fig.~\ref{eminfig} demonstrates radial profile of the minimum values of the specific energy of magnetized particles at stable orbits for the different values of MOG parameter. One can see from the Fig.~\ref{eminfig} with the increase of $\alpha$ parameter the value of the specific energy decreases and the distance where the particle energy is zero increases.

\begin{figure}
  \centering
   \includegraphics[width=0.98\linewidth]{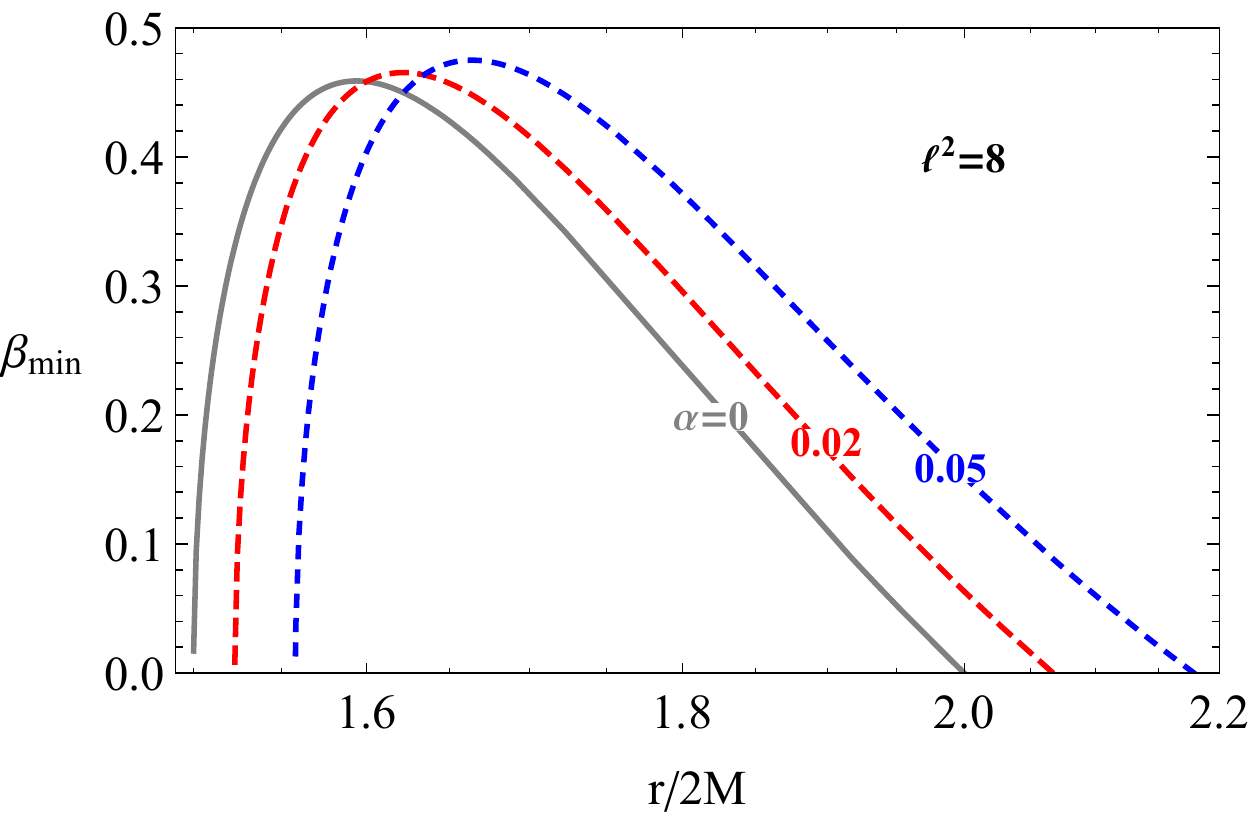}
    \caption{The radial dependence of minimal value of magnetic coupling parameter of the magnetized particle for the different values of the $\alpha$ parameter. The plots correpond to the values of the specific angular momentum $l^2=8$. \label{betaminfig}}
\end{figure}

Substituting (\ref{emin}) into (\ref{betafinal}) one may easily calculate the minimum value of the magnetic coupling parameter of magnetic particles for the given value of the specific energy  in the following form
\begin{eqnarray}\label{betamin}
\nonumber
\beta_{\rm min}&=&\frac{\sqrt{(\alpha +1)(2\alpha M-3r)M+r^2}}{ (\alpha +1)(\alpha M-2r)M+r^2}
\\
&\times & \Bigg\{\frac{l^2 \left[(\alpha +1)(2\alpha M-3r)M+r^2\right]}{(\alpha +1)(\alpha  M-r)M r}+r\Bigg\}.
\end{eqnarray}

Figure~\ref{betaminfig} illustrates the radial dependence of magnetic coupling parameter for different values of $\alpha$ parameter  for the fixed value of the specific angular momentum $l=2\sqrt{2}$. One can see that with the increase the value of the parameter, $\alpha$, the maximum value of the magnetic coupling parameter, $\beta$, and the distance where the magnetic coupling parameter, $\beta$, is zero increase.

Consider the upper limit for the angular momentum that the particle can be in stable circular orbits. The minimum value of the specific angular momentum corresponding to the minimum value of the magnetic coupling parameter can be found through the solution of the following condition $\partial \beta_{\rm min}/\partial r = 0$ with respect to $l^2$:
\begin{eqnarray}\label{lmineq}
\nonumber
l^2_{\rm min}&=&\frac{(\alpha +1)^2 M^2 r^2 (r-\alpha M)^2}{2 \alpha  (\alpha +1)M^2-3 (\alpha +1)M r+r^2}
\\
&&\left[\alpha  (\alpha +1) M^2-3 (\alpha +1) M r+2 r^2\right]^{-1}.
\end{eqnarray}

\begin{figure}[ht!]
  \centering
   \includegraphics[width=0.98\linewidth]{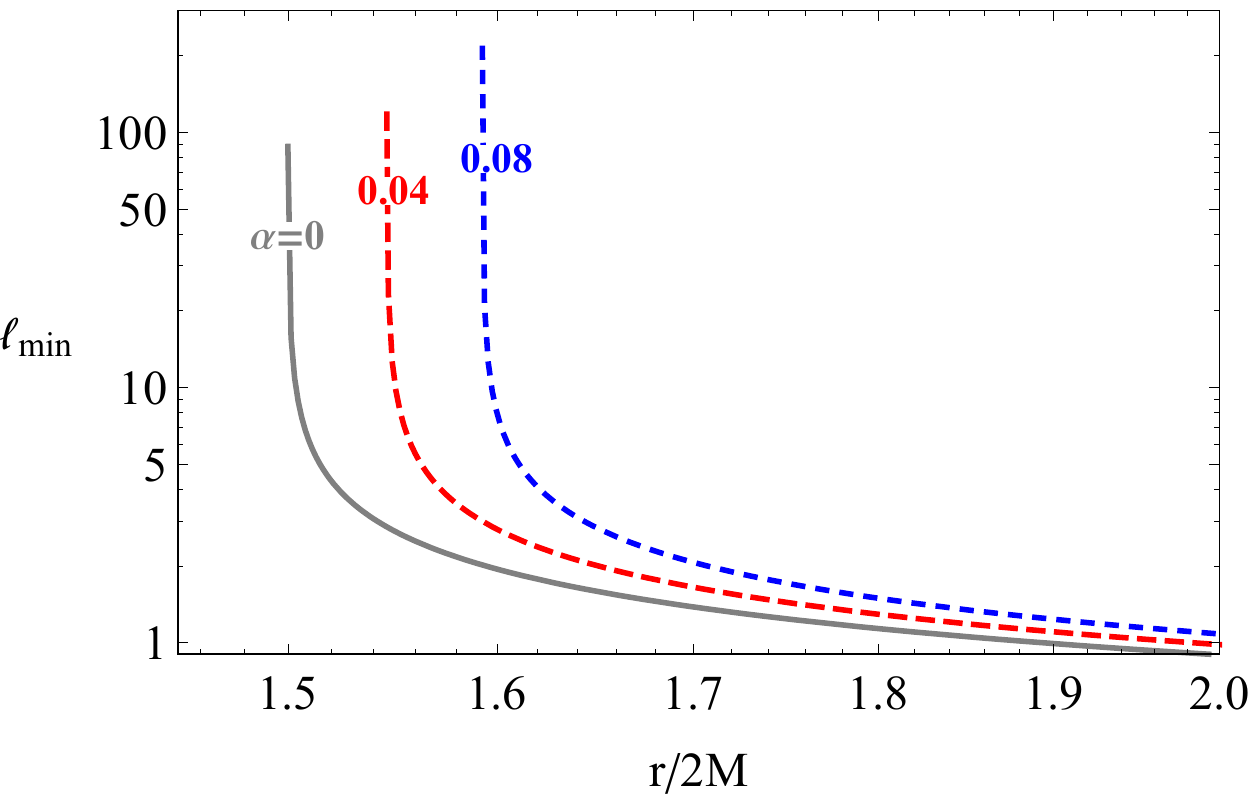}
    \caption{The radial profile of minimal value of specific angular momentum of the magnetized particle. \label{lminfig}}
\end{figure}

Figure~\ref{lminfig} shows the radial dependence of the minimum value of the specific angular momentum for the different values of the MOG parameter, $\alpha$. One can see from Fig.~\ref{lminfig} that the value of the specific angular momentum corresponding to a stable circular orbit and the radius where the angular momentum is maximum increase with the increase the parameter $\alpha$.

The extreme value of the magnetic coupling parameter $\beta$ can be found by omitting equation (\ref{lmineq}) into the equation (\ref{betamin}) in the following form
\begin{eqnarray}\label{betaext}
\beta_{\rm extr}=\frac{2 r \sqrt{2 \alpha  (\alpha +1)M^2-3 (\alpha +1) M r+r^2}}{\alpha  (\alpha +1) M^2-3 (\alpha +1) M r+2 r^2}\ .
\end{eqnarray}

\begin{figure}[ht!]
  \centering
   \includegraphics[width=0.98\linewidth]{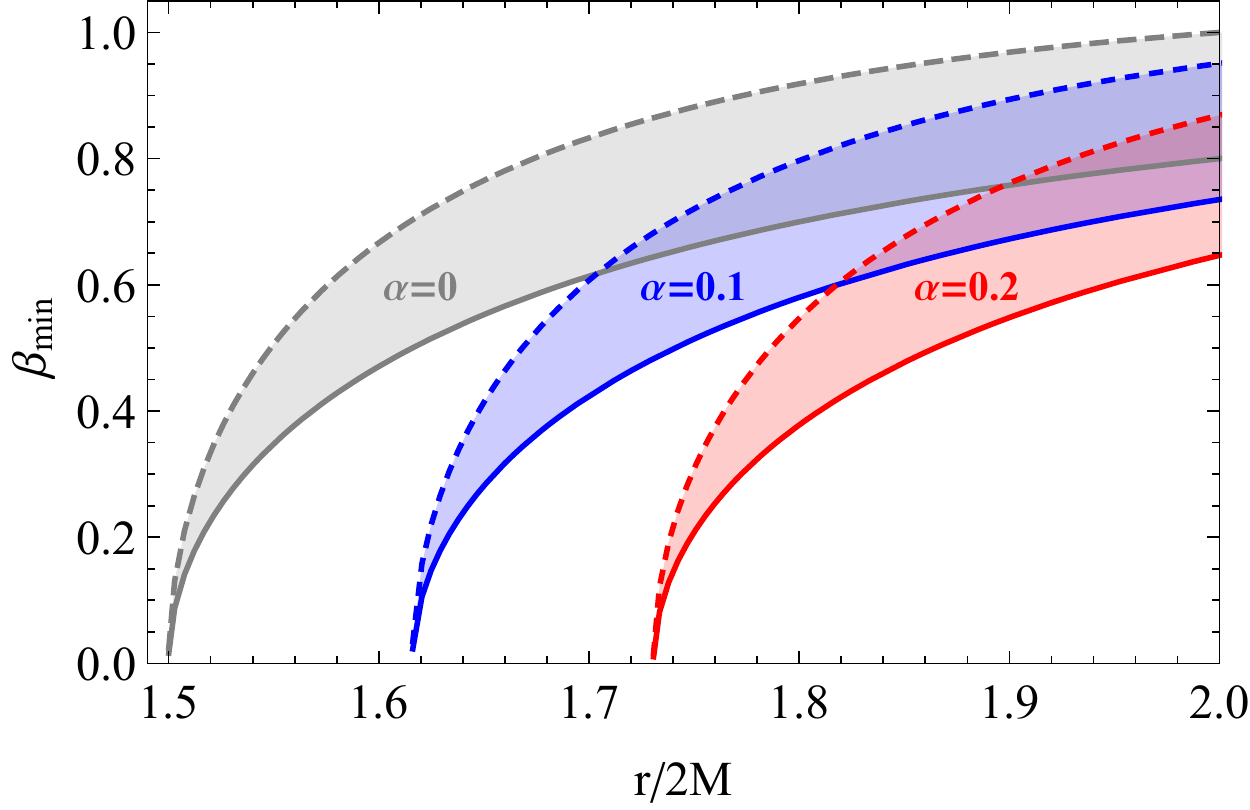}
    \caption{The radial profile  of minimal value of magnetic coupling parameter of the magnetized particle.     \label{betaextremfig}}
\end{figure}

Figure~\ref{betaextremfig} show the range of the magnetic coupling parameter corresponds to stable circular orbits. In the figure dashed lines correspond to the minimum value of the $\beta$ parameter at $l=0$ (freely falling magnetized particle) and solid ones correspond to the extreme value of the parameter $\beta$. Gray, light-blue and light-red colored areas correspond to the values of MOG parameter $\alpha=0$, $\alpha=0.1$ and $\alpha=0.2$, respectively. One can see from the Fig.~\ref{betaextremfig} that the distance where both the minimum and extreme values of the magnetic coupling parameter $\beta$ are zero shifts to the observer at infinity. However, the width of the area corresponding to the fixed value of the magnetic coupling parameter, $\beta$, does not depend on the parameter $\alpha$.

Thus, the extreme value of the parameter $\beta$ corresponds to maximum value of the critical stable circular orbits $r_{\rm max}$ and it can be found through the solution of the following equation with respect to  $r$
\begin{equation}\label{rmax}
\beta_{\rm ext}(r; \alpha)=\beta\ .
\end{equation}
The minimum value for the circular stable orbits can be found solving the following equation with respect to $r$,
\begin{equation}\label{rmin}
\beta_{\rm min}(r; \alpha)\, \vline\, _{l=0}=\beta\ .
\end{equation}

The distance between maximum and minimum radius of circular stable orbits $\Delta r= r_{\rm max}-r_{\rm min}$ give us the allowed area for  the stable orbits for a magnetized particle. That means the circular stable orbits of a magnetized particle with given $\beta$ are confined in the range $r_{\rm max}(\beta; \alpha)>r>r_{\rm min}(\beta; \alpha)$. However, one can see from the equations (\ref{betamin}) and (\ref{betaext}) it is quite complicated to obtain the analytic solutions of equations (\ref{rmax}) and (\ref{rmin}). We solve the equation numerically and present the results in a table form.

\begin{table*}[hbt] \begin{center}\begin{tabular}{|c| c| c| c| c| c| c| }\hline
$\alpha$ & $\beta=0.05$ & $\beta=0.1$ & $\beta=0.5$ & $\beta=0.8$ & $\beta=0.95$ & $\beta=1$ \\[1.5ex]\hline \hline
$0.01 $ & 0.00105 &  0,00423 & 0.1359 & 0.658274 & 2.1605 & $-$ \\[1.5ex]\hline
$0.05 $ & 0.00108 &  0,00437 & 0.1405 & 0.67895 & 2.6932 & $-$ \\[1.5ex]\hline
$0.1 $ & 0,0012 &  0,00453 & 0.1454 & 0.704716 & 2.7959 & $-$\\[1.5ex]\hline
$ 0.5 $ & 0.00144 & 0.00581 & 0.1868  & 0.908553 & 3.6096 & $-$ \\[1.5ex] \hline
$0.8$ & 0,00168 & 0.00676 & 0.1868 & 1.05976 & 4.2139 & $-$ \\[1.5ex]\hline
$1$ &  0.00183 & 0.00738 & 0.2375 & 1.1601 & 4.6153 & $-$ \\[1.5ex]\hline
$ 1.2 $ & 0.00198 & 0.00801 & 0.2581 & 1.26021 & 5.0158 & $-$ \\[1.5ex]\hline
$ 3 $ & 0.0033  & 0.01359 & 0.4394 & 2.15753 & 8,6058 & $-$\\[1.5ex]\hline
\end{tabular} \end{center}
\caption{\label{tab}  $\Delta r=r_{\rm max}-r_{\rm min}$ as function of $\beta$ and MOG parameter.}
 \end{table*}

The area of stable circular orbits of magnetized particles for the different values of MOG parameter is presented in Table~\ref{tab}  corresponding to the angular moment from $0$ to $l_{\rm min}$. One can see from the table the range $\Delta r$ increases as the increase of both parameters $\alpha$ and $\beta$, however, the effect of the $\beta$ parameter is stronger than effect of the parameter $\alpha$.

One may express the dependence of minimum values of the specific angular momentum on magnetic coupling parameter $\beta$ solving by the equation $\beta=\beta_{\rm min}$ with respect to the specific angular momentum $l$
\begin{eqnarray}\label{lbeta}
\nonumber
l_{\rm min}^2&=& \frac{(\alpha +1) M r (\alpha  M-r) }{\left(2 \alpha  (\alpha +1) M^2-3 (\alpha +1) M r+r^2\right)^{3/2}}
\\\nonumber
&\times & \Big\{\beta \left[r^2+\alpha  (\alpha +1)  M^2 \right]+\beta  r^2-2 (\alpha +1) \beta  M r
\\
&-& r \sqrt{2 \alpha  (\alpha +1) M^2-3 (\alpha +1) M r+r^2}\Big\}\ .
\end{eqnarray}

Now one can easily get the dependence of the minimum value of specific energy inserting Eq.~(\ref{lbeta}) into Eq.~(\ref{emin}):
\begin{eqnarray}
\nonumber
{\cal E}_{\rm min}^2&=&\frac{\left(\alpha  (\alpha +1) M^2-2 (\alpha +1) M r+r^2\right)^2}{r^3 \left(3 (\alpha +1) M r-2 \alpha  (\alpha +1) M^2-r^2\right)^{3/2}}
\\\nonumber
&\times & \Big\{ \beta  \left[r^2+\alpha  (\alpha +1)M^2\right]-2 (\alpha +1) \beta  M r
\\
&-&r \sqrt{2 \alpha  (\alpha +1) M^2-3 (\alpha +1) M r+r^2}
\Big\}\ .
\end{eqnarray}

\section{Particles collisions near the Schwarzschild-MOG black hole immersed in magnetic field\label{magaccel}}

In this section, we investigate particle acceleration processes near a Schwarzschild-MOG black hole by considering collisions of two magnetized particles in the presence of an external asymptotically uniform magnetic field. Here we will focus on the study of the effect of MOG parameter and external magnetic field to the center-of-mass energy of the colliding particles in-falling from infinity with specific (normalized to its mass) energies ${\cal E}_1$ and ${\cal E}_2$. The center of mass-energy of the two particles with the same mass $m$ can be found using the expression~\cite{Banados09}:
\begin{eqnarray}\label{ECMeq}
{\cal E}_{\rm cm}^2=\frac{E_{\rm cm}^2}{2m c^2}=1-g_{\alpha \beta}v_1^{\alpha}v_2^{\beta}\ ,
\end{eqnarray}
with $v_{1}^{\alpha}$ and $v_{2}^{\beta}$ are the 4-velocities of the colliding first and second particles, respectively.

We plan to consider collisions of a magnetized particle with charged and neutral particles in below.

\subsection{The case of two magnetized particles}

In this subsection we consider collisions of magnetized and magnetized particles at the equatorial plane where $\theta=\pi/2$, with the same initial energy. The components of four-velocity of the magnetized particle at  $\theta=\pi/2$ plane with $p_{\theta}=0$ has the following form:
\begin{eqnarray}\label{eqmotionm}
\nonumber
\dot{t}&=&\frac{{\cal E}}{f}\ ,
\\
\nonumber
\dot{r}^2&=&{\cal E}^2-f\left(1+\frac{l^2}{r^2}-\beta \right)\ ,
\\
\dot{\phi}&=&\frac{l}{r^2}\ .
\end{eqnarray}
The expression for ${\cal E}_{\rm cm}$ in this case can be defined using four-velocities of the particles at the equatorial plane and Eq.(\ref{ECMeq}) in the following form:

\begin{eqnarray}\label{cmmagmag}
\nonumber
{\cal E}_{\rm cm}^2&=&1+\frac{{\cal E}_1{\cal E}_2}{f}-\frac{l_1l_2}{r^2}-\\
\nonumber
&-&\sqrt{{\cal E}_1^2-f\left(1+\frac{l_1^2}{r^2}-\beta_1\right)}
\\
&\times & \sqrt{{\cal E}_2^2-f\left(1+\frac{l_2^2}{r^2}-\beta_2\right)}\ .
\end{eqnarray}
Now we will analyze the effect of the parameter $\alpha$ using Eq.(\ref{cmmagmag}) in a plot form
\begin{figure}
   \centering
  \includegraphics[width=0.98\linewidth]{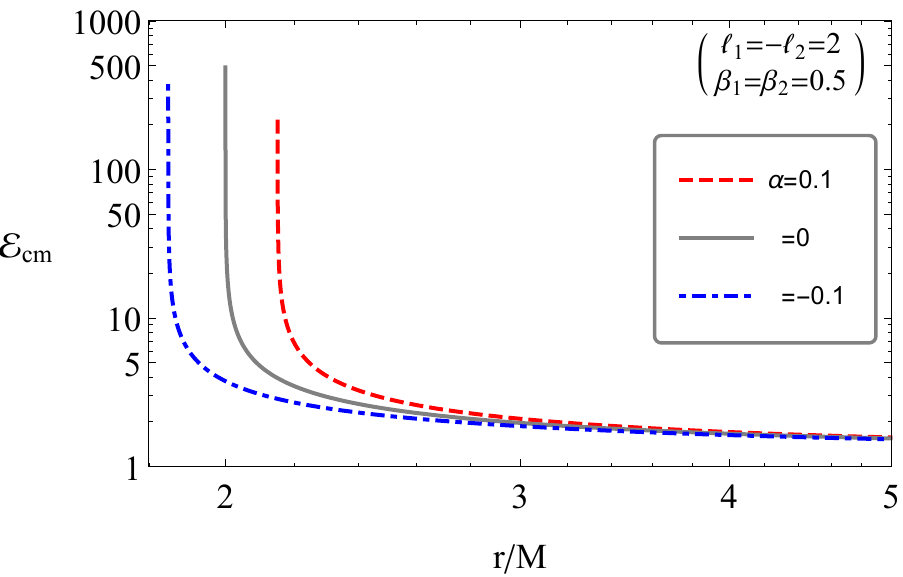}
	\caption{The radial profile of the ${\cal E}_{\rm cm}$ for the case of two magnetized particles with ${\cal E}_1={\cal E}_2=1$.  \label{centermm}}
\end{figure}

The radial profile of the ${\cal E}_{\rm cm}$ for two magnetized particles with the magnetized parameter $\beta_1=\beta_2=0.5$ for the different values of the MOG parameter $\alpha$ presented in Fig.~\ref{centermm} considering the collision with  $l_1=2, \, l_2=-2$. The Fig.~\ref{centermm} demonstrates that ${\cal E}_{\rm cm}$ increases with the increase of the MOG parameter.

\subsection{The case of magnetized and charged particles}

Now we will study the case  of magnetized and charged particles. The four-velocity of a charged particle can be found using the Lagrangian for the charged particle in curved space in the presence of electromagnetic field:
\begin{eqnarray}
{\cal L}=\frac{1}{2}mg_{\mu \nu}u^{\mu} u^{\nu}+q u^{\mu}A_{\mu} \ ,
\end{eqnarray}
where $q$ is the electric charge of the charged particle. The conservative quantities of the particle: energy and the angular momentum have the following form,
\begin{eqnarray}
E&=&
mg_{tt}\dot{t},
\\
L&=&
mg_{\phi \phi}\dot{\phi}+eA_{\phi},
\end{eqnarray}
and the four-velocity of the charged particle at the equatorial plane has the following components:
\begin{eqnarray}\label{eqmotionch}
\nonumber
\dot{t}&=&\frac{{\cal E}}{f}\ ,
\\
\nonumber
\dot{r}^2&=&{\cal E}^2-f\Big[1+\Big(\frac{l}{r}-\omega_{\rm B} r\Big)^2\Big]\ ,
\\
\dot{\phi}&=&\frac{l}{r^2}-\omega_{\rm B}\ , 
\end{eqnarray}
with $\omega_{\rm B}=eB/(2mc)$ being the  interaction parameter between the external magnetic field and the charged particle so called the cyclotron frequency responsible.

The expression for center-of-mass energy of magnetized and charged particles can be easily found  substituting Eqs.~(\ref{eqmotionch}), (\ref{eqmotionm}), in to (\ref{ECMeq}):
\begin{eqnarray}
\nonumber
{\cal E}_{\rm cm}^2&=&1+\frac{{\cal E}_1{\cal E}_2}{f}-\left(\frac{l_1}{r^2}-\omega_{\rm B}\right)l_2\\
\nonumber
&-&\sqrt{{\cal E}_1^2-f\Big[1+\Big(\frac{l_1}{r}-\omega_{\rm B} r\Big)^2\Big]}
\\
& \times & \sqrt{{\cal E}_2^2-f\Big(1+\frac{l_2^2}{r^2}-\beta \Big)}\ .
\end{eqnarray}

Here also we analyze the center of mass energy in a plot form.

\begin{figure}
    \centering
   \includegraphics[width=0.95\linewidth]{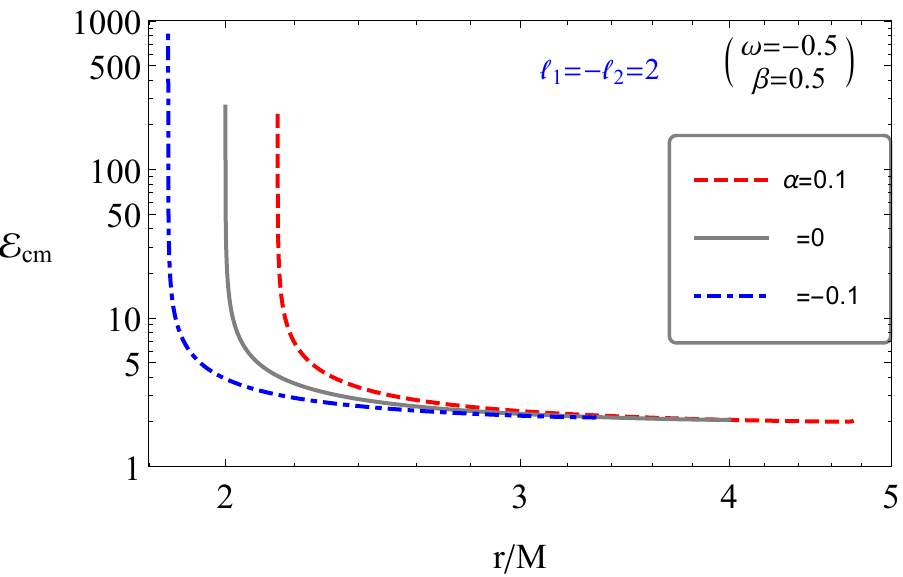}
   \includegraphics[width=0.95\linewidth]{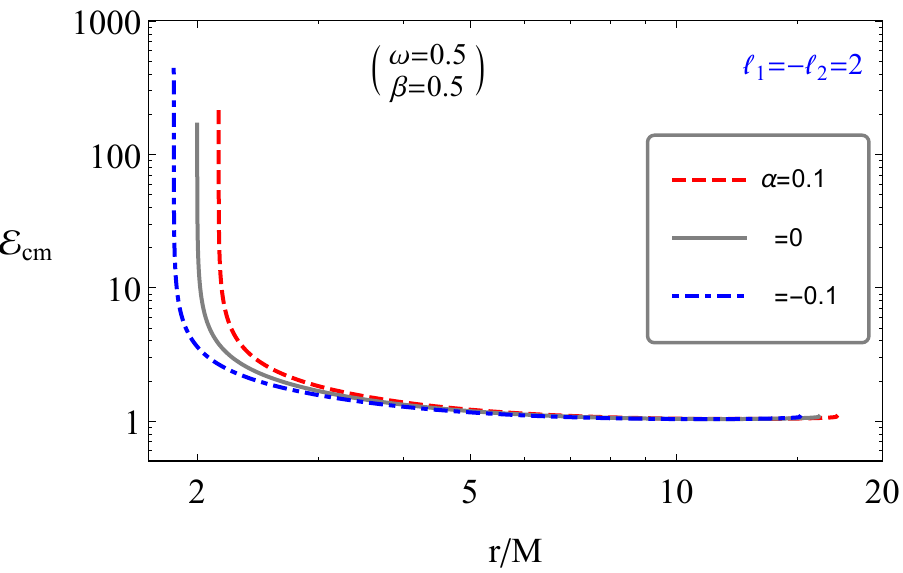}
   \caption{The radial profile of ${\cal E}_{\rm cm}$ for  charged and magnetized particles  with ${\cal E}_1={\cal E}_2=1$. Upper and down panels correspond to the cases of negative and positive charged particles, respectively.      \label{centermmq}}
\end{figure}

Figure~\ref{centermmq} shows the radial profile of ${\cal E}_{\rm cm}$ for the  magnetized and charged particles with the magnetic coupling parameter and cyclotron frequencies $\beta=\omega_{\rm B}=0.5$, around the Schwarzschild-MOG black hole for the different values of the parameter $\alpha$, considering the head-on collision with the values of specific angular momentum of the particles $l_1=2,\,l_2=-2$. One may explain the disappearance of the center-of-mass energy at large distances in both panels of Fig.~\ref{centermmq} by the repulsive Lorentz forces. 

\subsection{The case of magnetized and neutral particles}
In this subsection we will consider head-on collision of the magnetized and neutral particles. The equations of motion of neutral particles around Schwarzschild-MOG black hole can be written as:
\begin{eqnarray}\label{eqmotionn}
\nonumber
\dot{t}&=&\frac{{\cal E}}{f}\ ,
\\
\nonumber
\dot{r}^2&=&{\cal E}^2-f\Bigg(1+\frac{l^2}{r^2}\Bigg)\ ,
\\
\dot{\phi}&=&\frac{l}{r^2}\ .
\end{eqnarray}

The expression ${\cal E}_{\rm cm}$ for  the neutral and magnetized particles can be derived by substituting the Eqs.~(\ref{eqmotionm}), (\ref{eqmotionn}) in to (\ref{ECMeq}) and we have:
\begin{eqnarray}\nonumber
{\cal E}_{\rm cm}^2=1+\frac{{\cal E}_1{\cal E}_2}{f}-\frac{l_1l_2}{r^2}
&-&\sqrt{{\cal E}_1^2-f\left(1+\frac{l_1^2}{r^2}-\beta \right)}
\\
& \times & \sqrt{{\cal E}_2^2-f\left(1+\frac{l_2^2}{r^2}\right)}
\end{eqnarray}

\begin{figure}
    \centering
   \includegraphics[width=0.98\linewidth]{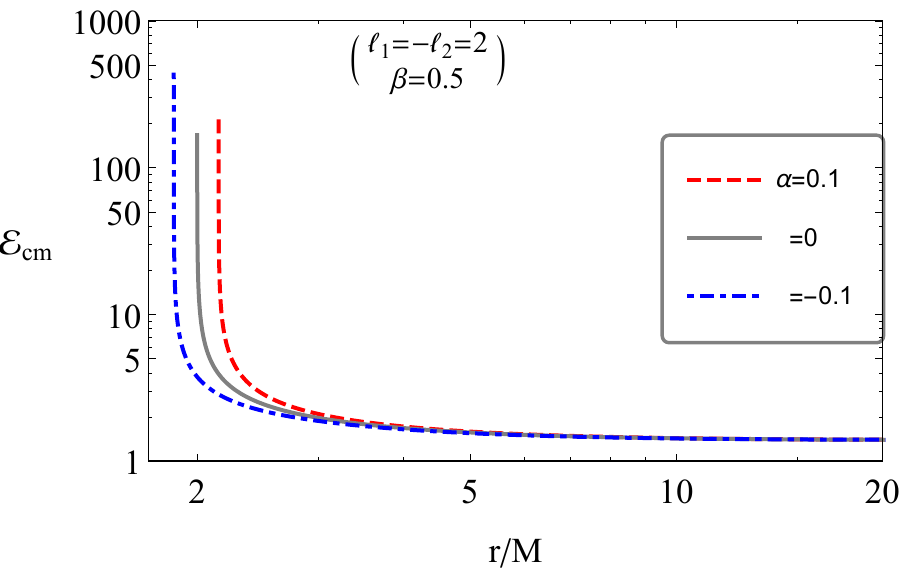}
      \caption{The radial profile of ${\cal E}_{\rm cm}$ for magnetized and neutral particles with ${\cal E}_1={\cal E}_2=1$. \label{centermn}}
\end{figure}

The radial profile of ${\cal E}_{\rm cm}$ for  neutral and magnetized particles with magnetic coupling parameter $\beta=0.5$, around the black hole in MOG for the different values of the parameter $\alpha$ shown in Fig.~\ref{centermn}. In this case, one may also see the center-of-mass energy of the collision increases (decreases) in the presence of positive (negative) MOG parameter $\alpha$.

\subsection{The case of two charged particles}

Here we will consider the energetic process from the collisions of two charged particles. The expression for the center of mass-energy of the two charged particle can be obtained inserting Eq.(\ref{eqmotionch}) into (\ref{ECMeq}) and we have
\begin{eqnarray}
\nonumber
{\cal E}_{cm}^2=1&+ &\frac{{\cal E}_1{\cal E}_2}{f}-r^2\left(\frac{l_1}{r^2}-\omega_{\rm B}^{(1)}\right)\left(\frac{l_2}{r^2}-\omega_{\rm B}^{(2)}\right)\\
\nonumber
&- &\sqrt{{\cal E}_1^2 -f\Big[1+\Big(\frac{l_1}{r}-\omega_{\rm B}^{(1)} r\Big)^2\Big]}
\\
&\times &\sqrt{{\cal E}_2^2-f\Big[1+\Big(\frac{l_2}{r}-\omega_{\rm B}^{(2)} r\Big)^2\Big]}\ .
\end{eqnarray}

\begin{figure*}
    \centering
   \includegraphics[width=0.48\linewidth]{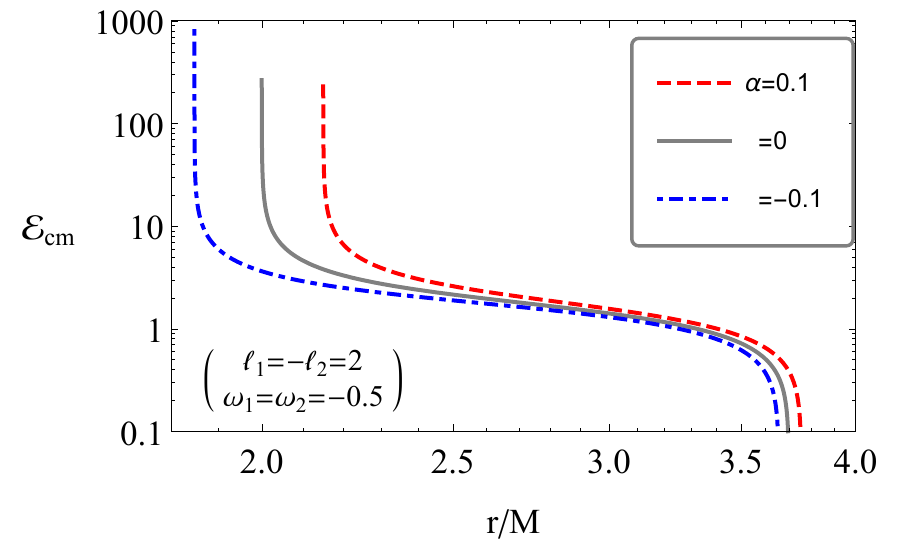}
   \includegraphics[width=0.48\linewidth]{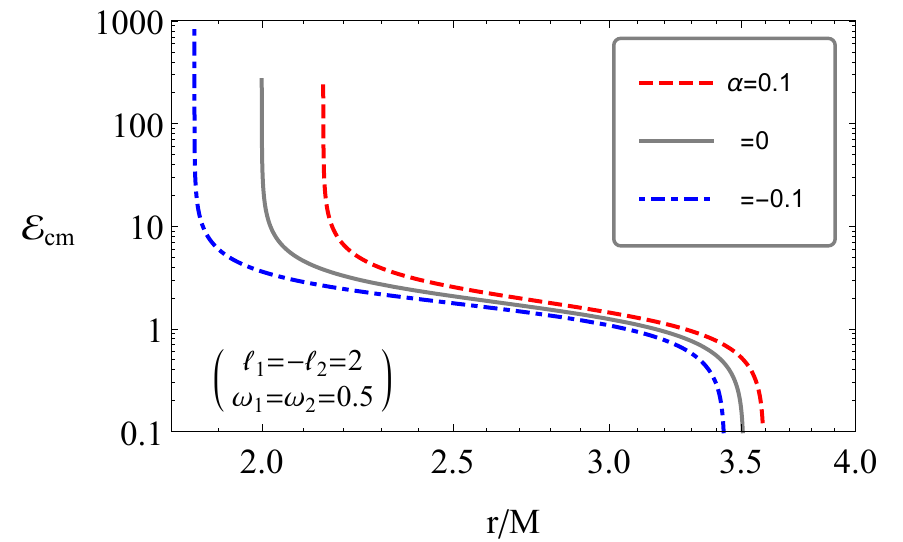}
   \includegraphics[width=0.48\linewidth]{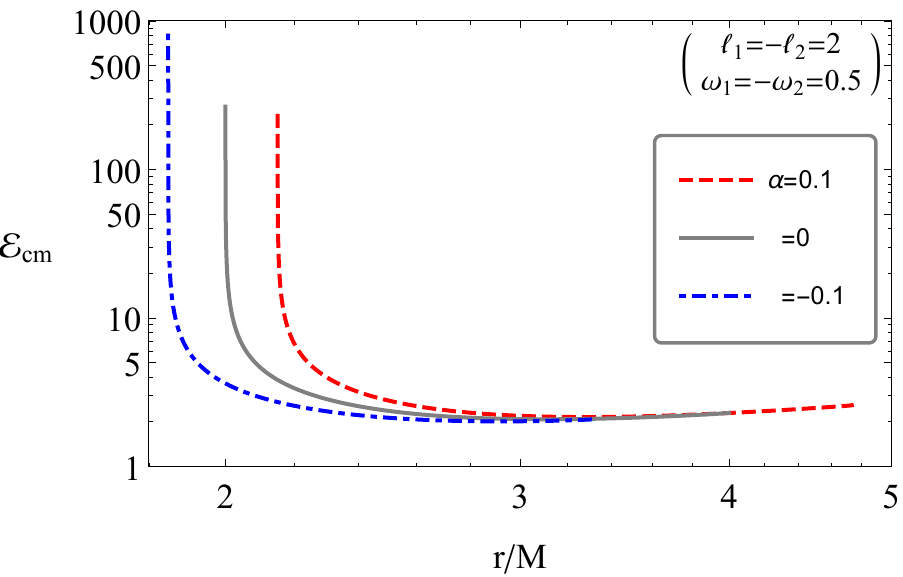}
   \caption{The radial profile of ${\cal E}_{\rm cm}$ for  charged particles with ${\cal E}_1={\cal E}_2=1$. Top-right, top left and bottom panels correspond to the cases of negative-negative, positive-positive and negative-positive charged particles, respectively.\label{centermmch}}
\end{figure*}

Theradial profile of ${\cal E}_{\rm cm}$ for charged particles near Schwarzschild-MOG black hole  is presented in Fig.~\ref{centermmch}. One can see from the figure that the center-of-mass energy increases (decreases) at the existence positive (negative) MOG parameter $\alpha$.  In the case of the collision of the negatively and positively charged particles, the distance where the center-of-mass energy disappears increases (decreases) in the presence positive (negative) MOG parameter $\alpha$ (bottom panel).

\section{Astrophysical applications \label{astroapll}}

As an astrophysical applications of the studies of magnetized particles around Schwarzchild-MOG black holes, we consider analysis of ISCO radius for the magnetized particle around the black hole in an external asymptotically uniform magnetic field and rotating Kerr black holes. In other words, we look for the answer to the question:  can magnetic interaction mimic the MOG and / or rotation parameters in the spectral fitting method implying necessity of additional methods in order to distinguish the Kerr black hole, the Schwarzschild-MOG black hole and presence of the external magnetic field.
The expression for the radius of ISCO of the test particles around Kerr BH is given by the relations~\cite{Bardeen72}
\begin{eqnarray}
r_{\rm ISCO}=3+Z_2-\sqrt{(3-Z_1) (3+Z_1+2 Z_2)}\, ,
 \end{eqnarray}
where
 \begin{eqnarray}\nonumber
Z_1 & = & 1+\left(\sqrt[3]{1-a}+\sqrt[3]{1+a}\right) \sqrt[3]{1-a^2}\, ,
\\\nonumber
Z_2 & = & \sqrt{3 a^2+Z_1^2}\, .
 \end{eqnarray}

Our aim is to study and compare ISCO radius of a magnetized particle around: (i) Kerr black hole, (ii) Schwarzschild-MOG black hole and Schwarzschild black hole immersed in an external asymptotically uniform magnetic field, corresponding the dimensionless values of rotation parameter $a$, $\alpha$ parameter and magnetic coupling parameter $\beta$ in the range $-1$ to $1$.
\begin{figure}[ht!]
  \centering
   \includegraphics[width=0.98\linewidth]{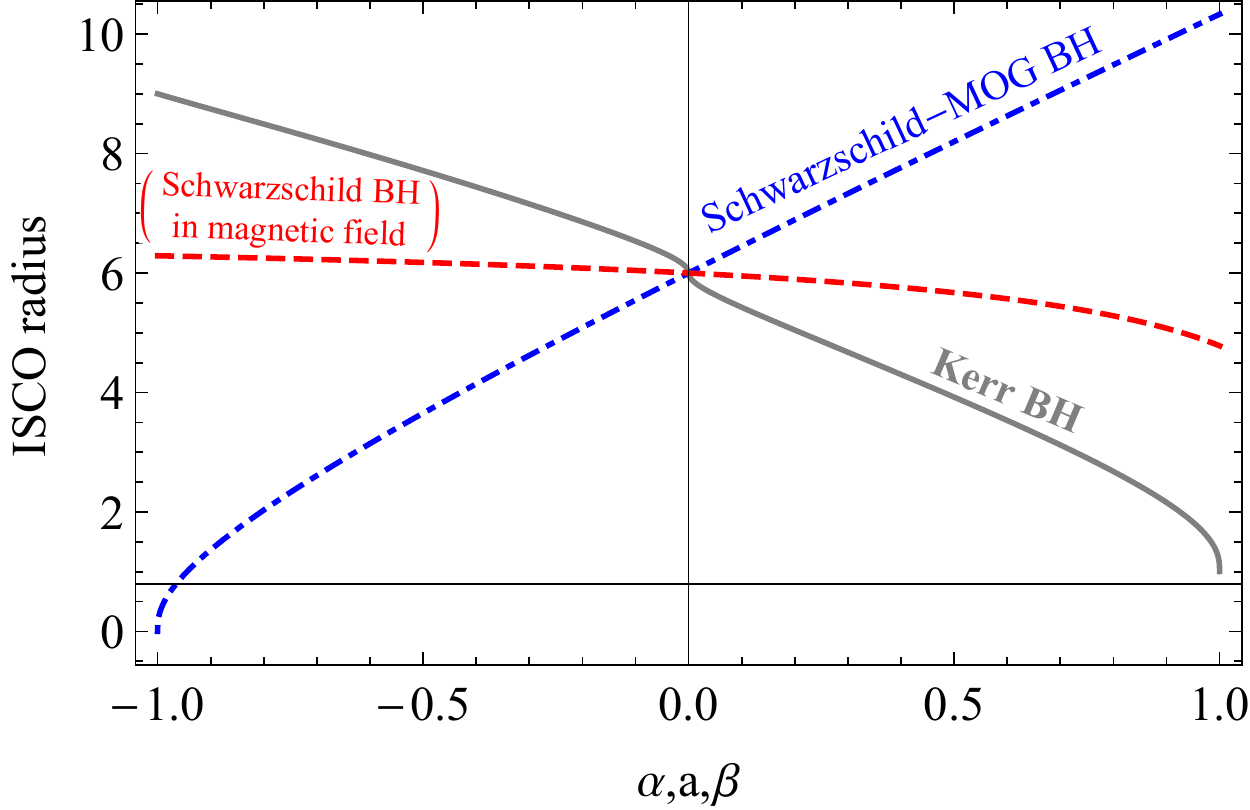}
    \caption{Dependence of ISCO radius on rotation (Gray colored solid line), MOG (blue colored dotdashed line) and magnetic coupling (red colored dashed line) parameters.}
   \label{iscoabeta}
\end{figure}

We show ISCO profiles of the magnetized particle around Schwarzschild black hole in the magnetic field (red-dashed line), Schwarzschild-MOG (dot-dashed blue line) and Kerr black holes (gray solid line) in Fig.~\ref{iscoabeta}. In this plot, the positive (negative) values of the magnetic coupling parameter  correspond to the same (opposite) direction of the external magnetic field with dipole momentum of particles and we consider negative spin parameter as co-rotation of the particle. One can see from the profiles that the parameters can give the same ISCO radius corresponding to their different values. This implies that the parameters can mimic each others in the observations of ISCO radius of magnetized particles.

\begin{figure}[ht!]
  \centering
   \includegraphics[width=0.98\linewidth]{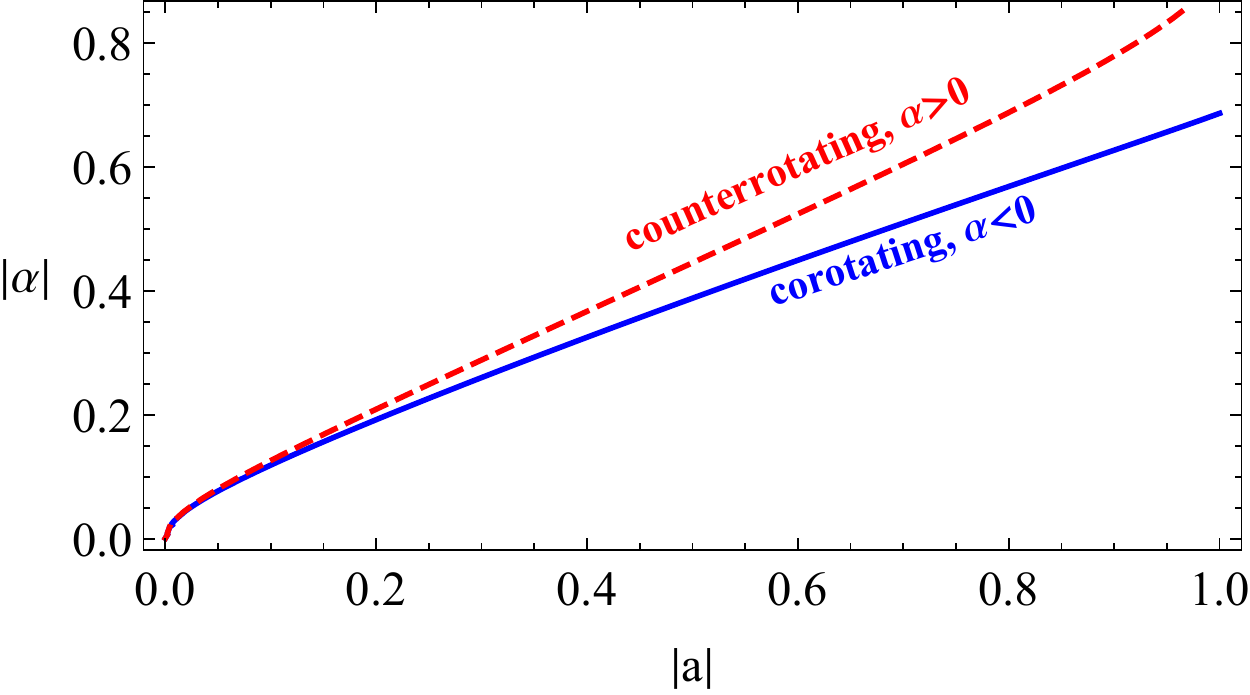}
    \caption{Relations between MOG and rotation parameters giving the same ISCO of the magnetized particles.\label{avsalpha}}
\end{figure}

First, we consider the particle motion around Kerr and Schwarzschild-MOG black holes for the same ISCO radius. Figure~\ref{avsalpha} shows the set of values of spin parameter of Kerr black hole and MOG parameter corresponding to the same value of the ISCO radius. One can see that positive (negative) values of the MOG parameter, $\alpha$ can mimic spin of Kerr black hole for the case of co-rotation (contour-rotation) of particles around the Kerr black hole. One may also see that the MOG parameter mimics at $\alpha \in (-0.7, 0.9)$ spin parameter of the black hole at the range $|a| \in (0,1)$.

\begin{figure}[ht!]
  \centering
  \includegraphics[width=0.98\linewidth]{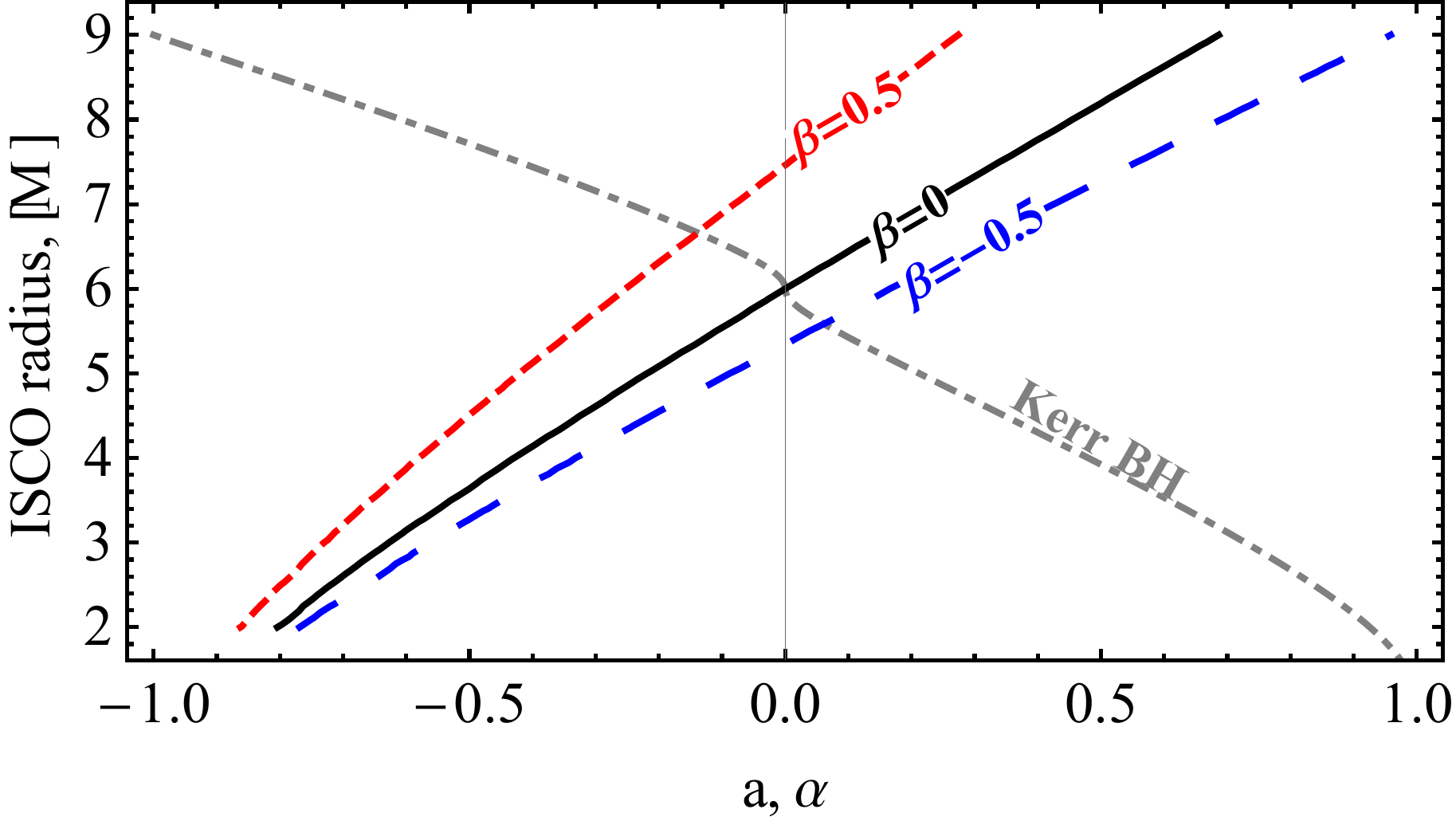}
    \caption{The dependence of the ISCO radius of test particles on spin of Kerr BH and MOG parameter. Red-dashed, blue-dashed, black-solid and gray-dotdashed lines are correspond to the ISCO of magnetized particles with magnetic coupling parameters $\beta=0.5$, $\beta=-0.5$, neutral particles around Schwarschild-MOG and Kerr black holes, respectively. \label{iscomagpart}}
\end{figure}

Now we consider magnetized particle motion around Scwarzchild-MOG and Kerr black hole. Figure~\ref{iscomagpart} shows the ISCO profiles of magnetized and neutral particles around Schwarzschild-MOG black hole in an external asymptotically uniform magnetic field and Kerr black hole. One can see from the figure that ISCO radius of particles around Kerr black hole $r_{\rm isco}=3M (7M)$ corresponds to the spin parameter $a=0.728533$ ($a=-0.395583$) which fits the value of MOG parameter at $\alpha \in (-0.730477, -0.562357)$ ($\alpha \in (-0.081132, 0.424942)$) for magnetized particles with the magnetic coupling parameters $\beta \in (-0.5, 0.5)$.

Now we will compare the ISCO radius of magnetized particles around the Kerr black hole and Schwarzschild black hole in the magnetic field and show how the magnetic field around Schwarzschild black hole can mimic the spin of Kerr black hole.

\begin{figure}[ht!]
  \centering
   \includegraphics[width=0.98\linewidth]{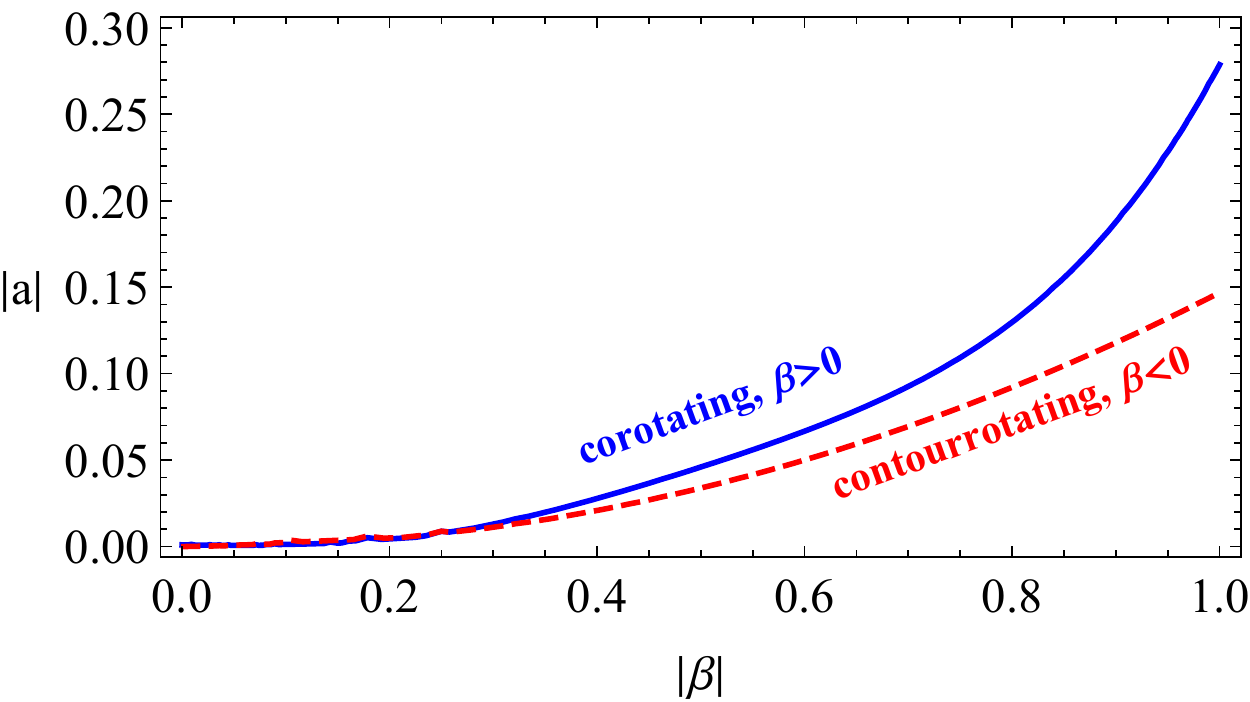}
    \caption{The similar plot as Fig.~\ref{avsalpha}, but for rotation and MOG parameters.}
   \label{abeta}
\end{figure}

In Fig.~\ref{abeta} we show the relations between the magnetic coupling parameter, $\beta$, and spin parameter, $a$, of Kerr black giving the same ISCO radius. One can see from the Fig.\ref{abeta} that positive (negative) values of $\beta$ parameter can mimic innermost co(contour)-rotating orbits of the particles around the Kerr black hole, giving the same radius of ISCO. One can see from Fig.\ref{abeta} that the magnetic coupling parameter can mimic spin of Kerr parameter $a \leq 0.15$ ($a \leq 0.28$) giving the same radius of innermost contour(co)-rotating orbits at the values of the magnetic coupling parameter $\beta \in (-1,1)$.

Finally, we will investigate the magnetized particle motion around Swarzschild-MOG black hole and Schwarzschild black hole in the magnetic field.

\begin{figure}[ht!]
  \centering
   \includegraphics[width=0.98\linewidth]{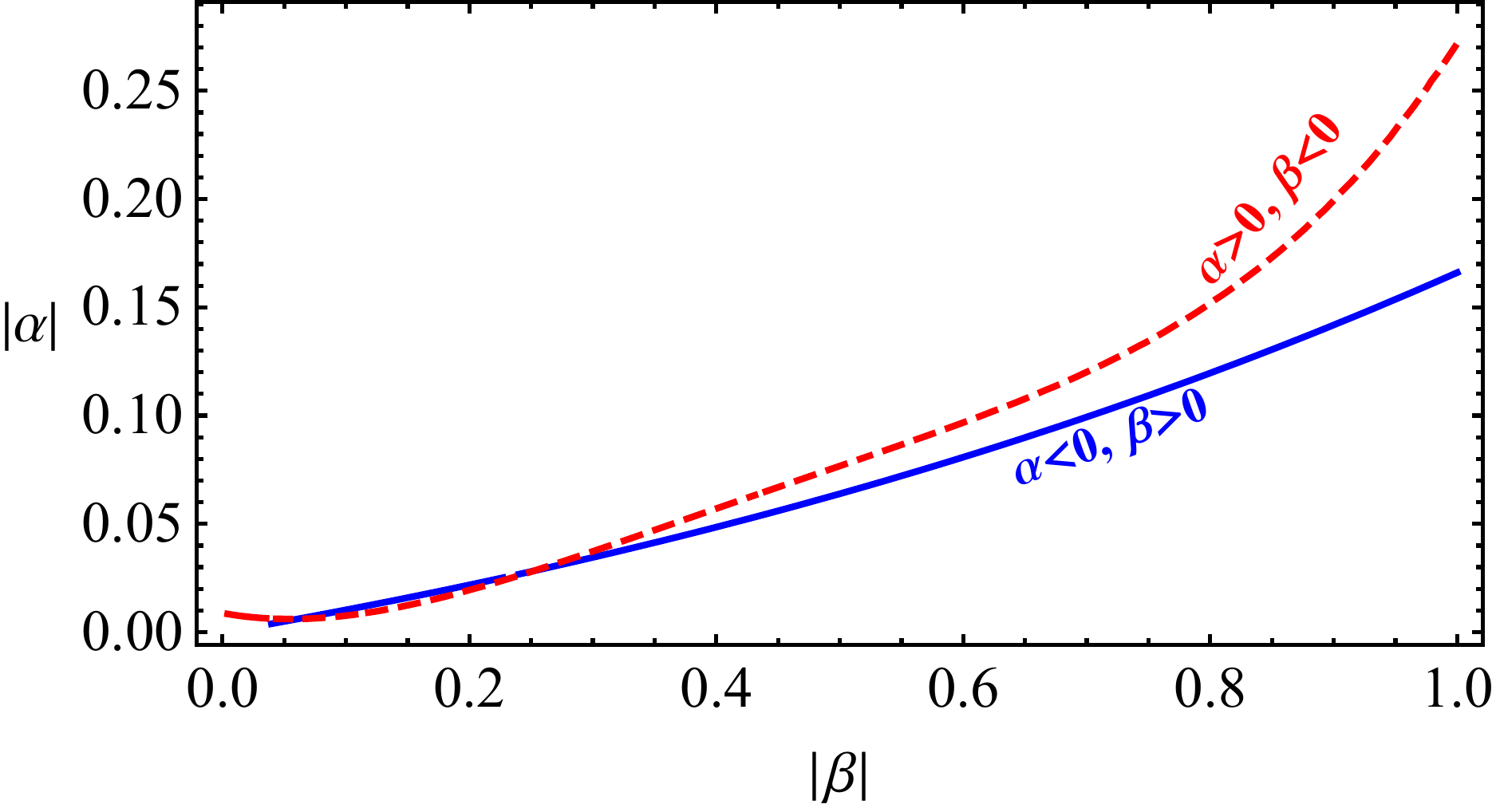}
    \caption{The similar plot as Fig.~\ref{abeta}, but for magnetic coupling and MOG parameters}
    \label{alpabeta}
\end{figure}

In Fig.~\ref{alpabeta}, we present relations of the magnetic coupling and the MOG parameters giving the same ISCO radius. From Fig.~\ref{iscomagpart} one can see that positive (negative) values of $\beta$ parameter can mimic negative (positive) values of the MOG parameter, $\alpha$, for the same radius of ISCO. One can see from the Fig.~\ref{alpabeta} that the parameter $\beta$ at ($\beta \in (-1,1)$) can mimic the MOG parameter in the range $\alpha \in (-0.17,0.28)$.

\section{Summary and Discussions\label{Summary}}

In this work, we have studied the motion of magnetized particles around Schwarzschil-MOG black hole immersed in an external asymptotically magnetic field. Analysis of circular orbits shows that the maximum value for the magnetic coupling parameter corresponding to the specific energy ${\cal E}=\sqrt{0.9}$ and angular momentum $l=\sqrt{10}$ decreases with the increase of MOG parameter. However, the maximum of the parameter $\beta$ does not change with the change of the parameter $\alpha$ for the values $l 2>12$. The minimum value of the magnetic coupling parameter near the Schwarzschild-MOG black hole increases with the increase of the $\alpha$ parameter and the minimum value disappears at $\alpha\geq 0.1$.

We have also studied the specific energy responsible for stable circular orbits and obtained that the energy decreases as the increase of parameter $\alpha$.
The studies of minimum and extreme values of $\beta$ magnetic coupling parameter show that orbits of the magnetized particles can not be stable at $\beta\geq1$.

Numerical calculations of the range between maximum and minimum stable orbits expand with increasing both $\alpha$ and $\beta$ parameters.

The investigations of collisions of magnetized particles have shown that the center-of-mass energy of the collisions increases with the increase of MOG-parameter, $\alpha$.

As an astrophysical application, we have investigated mimic values of magnetic coupling, MOG and spin parameters giving the same ISCO radius around Schwarzschild black hole in the magnetic field, Schwarz\-schild-MOG and Kerr black holes, respectively. Analysis of the studies have shown that \begin{itemize}

\item MOG parameter $\alpha \in (-0.7, 0.9)$ mimics spin parameter of black hole at the range $|a| \in (0,1)$;

\item the parameter $\beta$ at ($\beta \in (-1,1)$) can mimic the MOG parameter in the range $\alpha \in (-0.17,0.28)$;

\item and the parameter $\beta$ at ($\beta \in (-1,1)$) can mimic the MOG parameter in the range $\alpha \in (-0.17,0.28)$.

\end{itemize} Moreover, we have considered the similar studies of magnetized particles around Schwarzschild-MOG black hole immersed in an external asymptotically uniform magnetic field and Kerr black hole. Obtained that the magnetic coupling parameters $\beta \in (-0.5, 0.5)$ and the MOG parameter at $\alpha \in (-0.730477, -0.562357)$
($\alpha \in (-0.081132, 0.424942)$) can mimic the values of the spin parameter at $a=0.728533$ ($a=-0.395583$).

In our future works, in order to deeply understand the nature of modified gravity near the rotating black holes, we plan to extend the studies of magnetized particles motion around rotating Kerr-MOG black hole immersed in an external asymptotically uniform magnetic field.

\section{Acknowledgement}

AA is supported by PIFI postdoc fund by Chinese Academy of Sciences. This research is supported by Grants No. VA-FA-F-2-008 and No. MRB-AN-2019-29 of the Uzbekistan Ministry for Innovative Development.
This research is partially supported by an
Erasmus+ exchange grant between SU and NUUz.

\bibliographystyle{spphys}
\bibliography{gravreferences}

\end{document}